\documentstyle[psfig]{aa}
\newcommand{\gtap}{\mathrel{\hbox{\rlap{\lower.55ex \hbox {$\sim$}}
                   \kern-.3em \raise.4ex \hbox{$>$}}}}
\newcommand{\ltap}{\mathrel{\hbox{\rlap{\lower.55ex \hbox {$\sim$}}
                   \kern-.3em \raise.4ex \hbox{$<$}}}}

\begin{document}
\thesaurus{06(02.01.2; 08.09.2 YZ Cnc; 08.14.2)}
\title{X-ray observations through the outburst cycle of the dwarf nova YZ Cnc}
\author{Frank Verbunt\inst{1} \and Peter J. Wheatley\inst{2} \and
Janet A. Mattei\inst{3}
}
\offprints{Frank Verbunt email: verbunt@phys.uu.nl}

\institute{   Astronomical Institute,
              P.O.Box 80000, 3508 TA Utrecht, The Netherlands
         \and Department of Physics and Astronomy, University of Leicester,
              University Road, Leicester LE1 7RH, United Kingdom
         \and American Association of Variable Star Observers, 25 Birch Street,
              Cambridge MA\,02138-1205, United States of America
                }
\date{Received date; accepted date}   
\maketitle


\begin{abstract}
We have observed YZ Cnc at two day intervals from 6 to 24 April 1998,
covering two full outburst cycles.
The 0.1-2.4\,keV flux is lower during optical outburst than
in quiescence, and lowest at the end of the outburst.
The decline of the X-ray flux in the quiescent interval appears
to be in contrast to prediction of simple models for accretion-disk 
instabilities.
Variability on $\sim$hour time scales is present, but appears not
related to the orbital phase.
YZ Cnc was less luminous in X-rays during our 1998 observations than
in earlier ROSAT observations.

\end{abstract}

\section{Introduction}

Dwarf novae are named after their outbursts, during which their
luminosity at optical and ultraviolet wavelengths increases by
factors $\ltap 100$ (for a review see the monography on
Cataclysmic Variables by Warner 1995).\nocite{war95}
The outbursts are thought to be due to increased accretion 
onto the white dwarf.
Such an increase can be the consequence of increased transfer of
matter from the donor star to the accretion disk that surrounds the
white dwarf; alternatively an instability inside the accretion disk
could trigger higher accretion onto the white dwarf.
The latter model has been prominent in recent theoretical work, but
is not without difficulties (see reviews by e.g.\ Cannizzo 1993,
Verbunt 1991, Livio 1999).
\nocite{can93}\nocite{ver91}\nocite{liv99}

X-rays of dwarf novae arise from close to the white dwarf
and thus reflect the condition in the accretion disk close to the white dwarf
(see, e.g.,\ the review by Verbunt 1996).
For a comparison between different models, it is necessary to consider
the whole outburst cycle, including the quiescent interval (e.g.\ Pringle
et al.\ 1986).\nocite{pvw86}

YZ~Cnc is a member of the class of SU~UMa type dwarf novae, in which
short outbursts are occasionally interspersed by longer and
brighter superoutbursts.
Its orbital period is 0.0868(2)\,d (Shafter \&\ Hessman 1988); a more
accurate period of 0.086924(7)\,d is suggested by Van Paradijs et al.\ (1994).
\nocite{sh88}\nocite{pch+94}
YZ~Cnc is remarkable for the behaviour of its ultraviolet
resonance lines during its dwarf nova outbursts: during each orbit
the profiles of these lines change from almost pure emission 
to P Cygni profiles with deep absorption, and back again; these
changes are not accompanied by changes in the continuum
(Drew \&\ Verbunt 1988, Woods et al.\ 1992). \nocite{dv88}\nocite{wvc+92}
In X-rays YZ~Cnc has been studied with the Einstein satellite 
(C\'ordova \&\ Mason 1984, Eracleous et al.\ 1991), 
with EXOSAT (van der Woerd 1987),
and with the ROSAT PSPC during the ROSAT All Sky Survey and in
subsequent pointings (Verbunt et al. 1997, van Teeseling \&\ Verbunt 1994).
\nocite{cm84}\nocite{ehp91} \nocite{woe87} \nocite{vbrp97}\nocite{tv94}

In this paper we report on a ROSAT campaign intended to determine
the X-ray fluxes throughout the outburst cycle of YZ~Cnc; and also
to determine whether the orbital variation in the ultraviolet lines is
accompanied by orbital variation in the X-ray flux.
In Sect.\ 2 we describe the observations and data analysis,
the results and their interpretation are given in Sect.\ 3; 
comparison with earlier X-ray observations is made in Sect.\ 4,
and the implications for the models of dwarfs nova outbursts are
discussed in Sect.\ 5.

\section{Observations and data reduction}

\begin{table}
\caption{Log of the ROSAT HRI observations of YZ Cnc in April 1998, and
one PSPC observation obtained on May 1, 1994.
For each observation we give the UT at start and end of the exposures and the 
effective exposure time; as well as the observed countrate (with the 
1-$\sigma$ error in the last two decimals).
\label{tablog}}
\begin{tabular}{llrl}
date & exposure start \&\ end & $t_{\rm exp}$ & countrate \\
      &                       & (s)           & (cts/s) \\
1994 May 1    & 2449473.614-73.655 &  1369 & 0.403(18)$^a$ \\
1998 April 6  & 2450909.935-10.090 &  6708 & 0.1121(41) \\
1998 April 8  & 2450911.921-12.152 & 10220 & 0.0208(15) \\
1998 April 10 & 2450913.912-14.142 &  9417 & 0.0124(12) \\
1998 April 12 & 2450915.905-16.132 &  9053 & 0.0971(33) \\
1998 April 14 & 2450917.893-18.120 &  9425 & 0.0842(30) \\
1998 April 16 & 2450919.813-20.041 &  9427 & 0.0236(16) \\
1998 April 18 & 2450921.802-22.031 &  9925 & 0.0156(13) \\
1998 April 20 & 2450923.790-24.020 &  9755 & 0.0990(32) \\
1998 April 22 & 2450925.713-25.942 &  9245 & 0.0866(31) \\
1998 April 24 & 2450927.702-27.929 &  8540 & 0.0191(15) \\
\end{tabular}

$^a$PSPC countrate in channels 11-235.
\end{table}

All 1998 observations were obtained with the ROSAT X-ray telescope
(Tr\"umper et al.\ 1991) in combination with the high-resolution imager
(HRI, David et al.\ 1995).\nocite{tha+91}\nocite{dhkz95}
The log of the observations is given in Table~1.

The data reduction was done with the Extended Scientific Analysis System
(Zimmermann et al.\ 1996).\nocite{zbb+96}
YZ Cnc was detected in every pointing, the countrate was determined
by applying a maximum-likelihood technique which compares the observed
photon distribution with the point spread function of the HRI
(Cruddace et al.\ 1988).\nocite{chs88}
The resulting countrates are given in Table~1. YZ Cnc is much brighter
than the background countrate; the errors in the countrate are therefore
dominated by Poisson statistics on the detected number of source counts.

\begin{figure*}
\centerline{\psfig{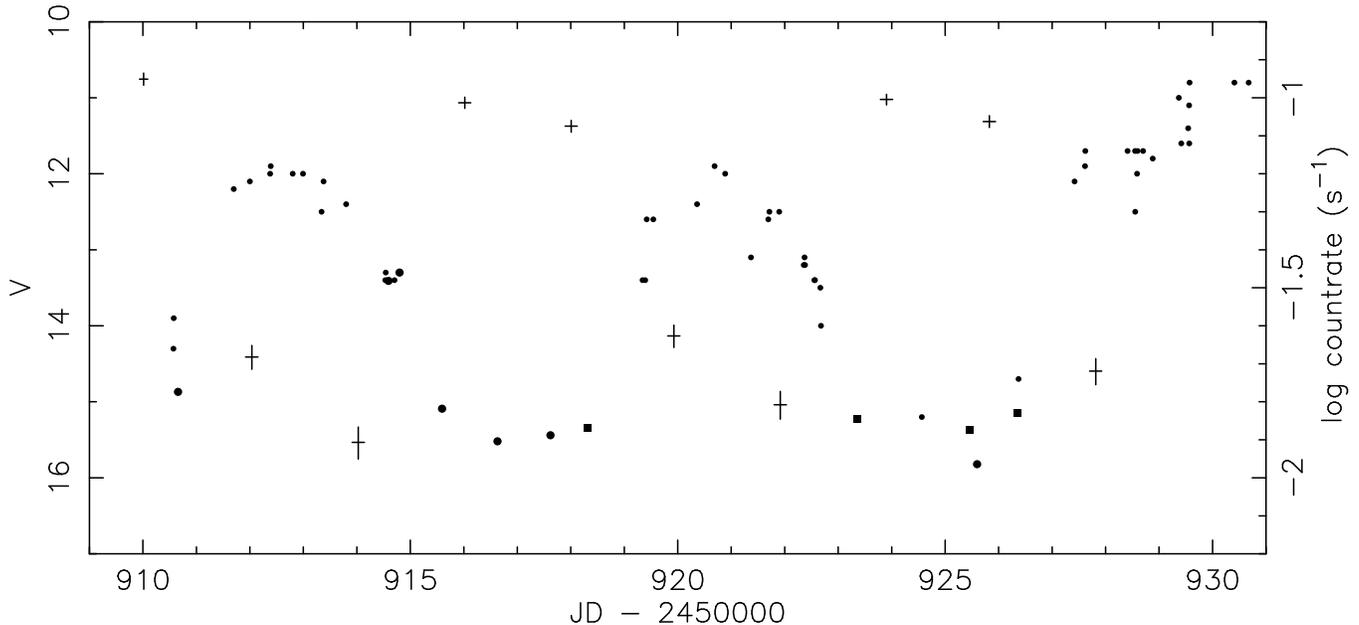} {\hfil}}
\caption{Optical and X-ray lightcurve of YZ Cnc in April 1998. The optical
data (magnitude scale on the left) have been obtained by the American 
Association of Variable Star Observers: 
small dots indicate estimates made by eye; the large symbols indicate CCD
measurements. The HRI countrates (logarithmic scale on the right) are shown as
crosses, the horizontal and vertical lengths of which indicate the 
observation period and 1-$\sigma$ error range, respectively. 
\label{curve}}
\end{figure*}

\section{Results and interpretation}

We investigate the variation of the X-ray flux of YZ Cnc through the
outburst cycle, and also on the orbital timescale.

\subsection{Outburst cycle}

In Figure~1 we show the optical lightcurve of YZ~Cnc from April 5 to 26, 1998,
as determined by the American Association of Variable Star
Observers, together with the HRI countrates listed in Table~1.
The optical lightcurve shows maxima of ordinary outbursts occurring
on JD\,2450912 and JD\,2450921, and a superoutburst maximum near
JD\,2450930.

In comparing the X-ray countrates with the optical lightcurves we remark
on three features of Fig.~1.
First, the HRI countrates are lower during the optical outbursts
than in the quiescent intervals.
Second, in both quiescent intervals that we cover, the countrate is lower
in the later observation.
Third, in both outbursts that we cover, the countrate is lower
in the later observation.

The distribution of the photons over the energy channels of the HRI
is the same for the observations taking during outburst as for those
taken during quiescence.
Comparison of the distributions obtained for the first observations
during outburst (i.e.\ those of April 8 and 16) with those obtained
during the later outburst observations (of April 10 and 18) suggests that
the decrease in X-rays  is marginally less at the lower energies, i.e.\ that
the spectrum becomes slightly softer as the optical outburst proceeds.
The significance of this softening is marginal, but it suffices to
show that the decrease of the X-ray flux cannot be due to the
disappearance of an ultra-soft component.

In accordance with these findings, we interpret the change in HRI countrates
during the outburst cycle as a change mainly in the amount of gas that emits
$~$\,keV photons.
This amount drops gradually during quiescence, more dramatically
in the beginning of an outburst, and gradually again as the outburst
proceeds.

\subsection{Short-term variability}

We have searched for short-term variability by dividing the individual
observations in smaller intervals.
For the outburst data we determine the average countrate during each ROSAT
orbit; for the higher countrate during quiescence we use bins of 256\,s.
Figure~2 show the resulting lightcurves.
Significant variation is present both during outburst and during
quiescence. 
During early ordinary outburst (April 8,16) the variation appears dominated 
by a long-term decline. No orbital variation is apparent in any of the
outburst data.
During quiescence, the flux level at a given orbital phase varies as much as 
the overall variation. 
We have folded the variation on the orbital period of 0.086924\,d,
and find no significant variation on the orbital period, in quiescence 
or in outburst. 
Any orbital variation is less than the irregular variations seen
in Fig.~2.

\begin{figure}
\centerline{\psfig{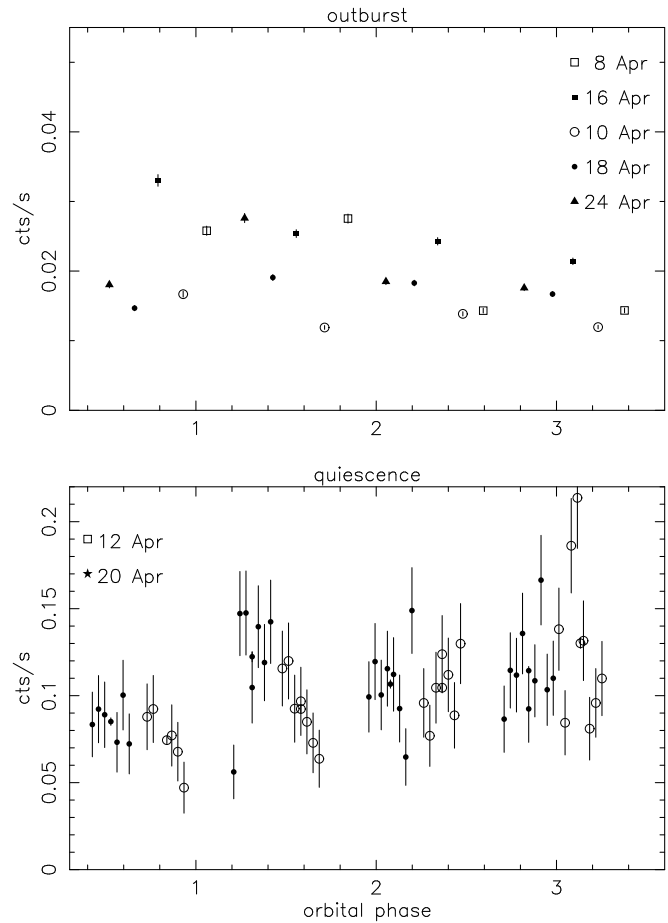} {\hfil}}
\caption{X-ray variability of YZ Cnc during outburst and in quiescence.
Data are shown for all observations obtained during outburst, showing one 
point for each ROSAT orbit, and for two observations in early quiescence,
showing average countrates in 256\,s bins. 
1-$\sigma$ error are indicated. 
The time axis is in units of orbital phase, for the period
according to Van Paradijs et al.\ (1994), but with arbitrary zero point.
\label{xcurv}}
\end{figure}

\section{Comparison with previous X-ray observations}

\subsection{Previous observations of YZ Cnc}

To compare our observations with previous ROSAT PSPC observations,
we note that for a 2-3\,keV thermal spectrum as found for YZ Cnc
by Van Teeseling \&\ Verbunt (1994) the ROSAT PSPC (channels 50-201) countrate
is similar to the Einstein IPC countrate, and about twice the
ROSAT HRI countrate.
From the results listed in Table~1, we therefore expect countrates in the
ROSAT PSPC (ch.\,50-201) or Einstein IPC of 0.22-0.17 cts/s in quiescence
and of 0.047-0.025 cts/s during outburst.

The All-Sky Survey observation was obtained from 10 to 12 October 1990,
i.e.\ during the outburst which peaked on October 10 (Bortle 1990).
The countrate (in PSPC channels 52-201) is about 0.1\,cts/s (Verbunt et 
al.\ 1997), and does not vary significantly.
The pointed observations with the ROSAT PSPC gave countrates 
(in channels 50-201) of 0.4 cts/s on 3 April 1991 immediately before an 
optical outburst maximum, and of 0.27 cts/s on 7-11 October 1993
in quiescence (van Teeseling \&\ Verbunt 1994).
We have analyzed a previously unpublished ROSAT PSPC observation, obtained
on 1 May 1994. The countrate (ch.\, 50-201) is 0.249(14) cts/s, marginally 
lower than the 1993 countrate.
All these countrates are significantly higher than the corresponding
ones during outburst and in quiescence in April 1998,
and indicate long-term variability in both quiescent and outburst
X-ray fluxes of YZ Cnc.

An Einstein IPC countrate of about 0.04\,cts/s was observed on 8 April 1979
(C\'ordova \&\ Mason 1984), and has been hitherto interpreted as obtained 
during quiescence. 
The Einstein countrate corresponds to the outburst level of April 1998.
Outbursts of YZ Cnc were observed by the AAVSO peaking on March 8, 18 and 
28 and on April 15 and 23 in 1979 (Bortle 1979).
AAVSO measurements of YZ Cnc in March and April 1979 are shown
in Fig.~3.
The quiescent interval separating the March 28 and April 15 outbursts was 
longer than the intervals preceding and following it.
A single, uncertain measurement obtained close in time to the Einstein
observation suggests that YZ Cnc was brighter than its quiescent level.
We suggest that the Einstein observation was obtained during an
outburst peaking close to 8 April 1979, which was missed by the optical
observers.

\begin{figure}
\centerline{\psfig{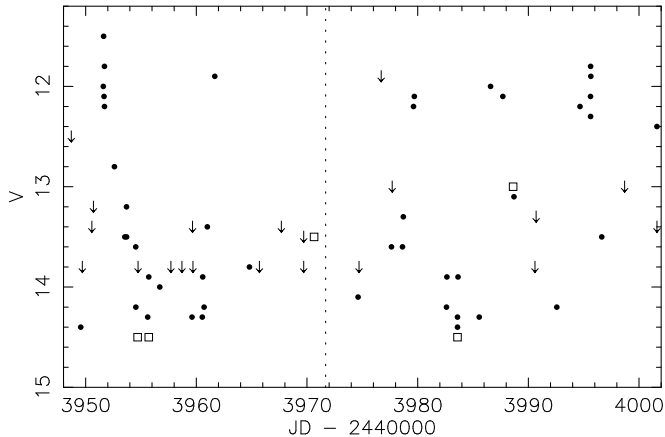} {\hfil}}
\caption{Visual magnitude of YZ Cnc between March 16 and May 8 1979,
as measured by the AAVSO. A magnitude estimate is shown
as $\bullet$, an uncertain estimate as $\Box$, and an upper limit as
$\downarrow$. The vertical dotted line indicates the time of the
Einstein observation.
\label{aaein}}
\end{figure}

The countrates observed with the EXOSAT Low Energy detector (with 3000 Lexan
filter) between 27 October and 13 November 1983 were about 0.01 cts/s
during quiescence, and a factor 4 lower during outbursts (Van der Woerd
1987). 
For an assumed column in the range $N_{\rm H}=10^{19}-10^{20}{\rm cm}^{-2}$,
an HRI countrate of 0.1\,cts/s predicts an countrate for the EXOSAT LE
(3000 Lexan) in the range $0.02-0.01$\,cts/s.
Depending on the assumed column, the EXOSAT observations are thus
compatible both with the higher X-ray luminosity as observed with the
ROSAT PSPC observations, and with the somewhat lower luminosity
of our ROSAT HRI observations.

\subsection{Orbital variability}

Our finding that the observed X-ray flux of YZ~Cnc varies on short
time scale unrelated to the orbital phase, is in accordance with
similar findings by Van der Woerd (1987) in his analysis of the EXOSAT
data.
Any explanation of the marked change in the profiles of the ultraviolet 
resonance lines in YZ Cnc in terms of a variable absorption column between
the ultraviolet continuum source and Earth, must be compatible with a
much less marked variation in the X-ray flux.
For a cold gas with cosmic abundances, an upper limit in X-ray
variability on orbital time scales of $\ltap10\,$\%\ 
(see the April 18 data in Fig.\,2) corresponds to
an upper limit in the column of $N_{\rm H}\ltap10^{20}\,{\rm cm}^{-2}$.
The upper limit to the column in the more realistic case of the
complicated ionization structure in the wind of a dwarf nova can be
determined only in a detailed model of this structure (see e.g.\ the
review by Drew 1997 and references therein).\nocite{dre97}

\subsection{Observations of other dwarf novae}

Various other dwarf nova outbursts have been covered in X-rays.
A general description applies to the EUV/X-ray lightcurves of VW Hyi, 
Z Cam and SS Cyg observed so far: a hard component is more luminous in 
quiescence, and less 
luminous but constant during outburst, whereas a soft component is brighter 
during outburst than in quiescence, and decreases rapidly after reaching 
its maximum early in the outburst (Wheatley et al.\ 1996a,b,
Ponman et al.\ 1995). \nocite{wvb+96}\nocite{wtw+96}\nocite{pbd+95}
The situation is different for the dwarf nova outbursts of U Gem, where
both soft (0.15-0.5\,keV) and hard (2-10\,keV) X-ray fluxes
are higher during outburst than in quiescence; and both components
appear to decrease faster during the outburst than the optical
flux (Mason et al.\ 1978, Swank et al.\ 1978).\nocite{mlcb78}\nocite{sbh+78}
The EUVE observations of U Gem and SS Cyg show that the soft component is not 
alltogether optically thick (Mauche et al.\ 1995, Long et al.\ 1996).
\nocite{mrm95}\nocite{lmr+96}

The soft component has a different temperature in each of the above 
dwarf novae, but in most cases the HRI band is dominated by the hard 
component. (The exception is SS Cyg, whose soft component has a
relatively high characteristic temperature; Van Teeseling 1997.)\nocite{tee97}
This supports our conclusion in Sect.\, 3.1 that the hard component
is responsible for the observed decrease during outburst  of the X-ray flux
in YZ Cnc.

As regards the quiescent interval between outbursts,
we are aware of only one other system that has been observed
throughout several full outburst cycles, viz.\ VW Hyi.
EXOSAT observations of this system showed a decrease of the flux 
in the 0.05-1.5\,keV energy range during each of three covered
quiescent intervals (van der Woerd \&\ Heise 1987, Pringle et al.\ 1987),
in accordance with our findings for YZ~Cnc.
\nocite{wh87}\nocite{pbh+87}

\section{Implications for outburst models}

In Sect.\,3.1 we concluded that the changes in X-ray flux of YZ Cnc that we
observe through the outburst cycle are due mainly to changes in the
amount of X-ray emitting gas. 
We will compare our results with the predictions of the two classes
of models of dwarf nova outbursts, the mass transfer instability and
the disk instability.

Both models can explain a lower X-ray flux during the outburst 
as a consequence of a transition of an optically thin, very hot boundary 
layer during quiescence into a rather less hot, optically thick gas 
during outburst when the accretion rate onto the white dwarf is high
(Pringle \&\ Savonije 1979).\nocite{ps79}
The observations of dwarf novae in outburst indicate that not all
the X-ray emitting gas disappears during the outburst, but that
some of it remains, at much the same temperature as during quiescence.
It would be tempting to locate this remaining component away from the
disk, e.g.\ in a white dwarf or disk corona, if it is to escape becoming
optically thick with the increased accretion rate.
However, it then has to be explained why this component has a spectrum
very similar to the component in the disk, and why it changes
at all during outburst.
A location of the X-ray emitting gas separate from the optically thick
accretion disk could also be an ingredient in understanding how
the small variation in X-rays is compatible with a large absorbing
column required to explain the strong variations in the ultraviolet
resonance lines.

The mass transfer instability model is not sufficiently developed
to predict the mass transfer as a function of time, and thus to
predict the evolution of the X-ray flux, during outburst.
However, the model does predict a continued decrease of the accretion rate
onto the white dwarf during quiescence, perhaps levelling off to a
constant level in long quiescent intervals when the disk reaches 
equilibrium with the lower mass inflow rate at its outer edge.
A continuing decrease of the X-ray flux in quiescence, as observed
for YZ Cnc and various other dwarf novae, is thus in accordance with
the transfer instability model (for model accretion
rates onto the white dwarf during the outburst cycle, see e.g.\ Pringle
et al.\ 1986).

The disk instability model in its simple form predicts a gradual
increase of the accretion rate onto the white dwarf during quiescence,
and there therefore a gradual increase in the optical and ultraviolet flux.
An increased accretion rate through an optically thin disk also predicts
an increase in the X-ray flux, contrary to our observations of YZ Cnc.

The rise of the ultraviolet flux in quiescence predicted by the disk
instability model is contrary to observations of several dwarf novae,
in particular the eclipsing system Z~Cha, but also VW~Hyi and WX~Hyi
(Van Amerongen et al.\ 1990, Verbunt et al.\ 1987, Hassall et al.\ 1985).
\nocite{akp90}\nocite{vhp+87}\nocite{hpv85}
Szkody et al.\ (1991) investigated the evolution of the ultraviolet flux of 
many dwarf novae in quiescence, and did not find a
single case where the ultraviolet flux increases {\it when measured in a single
interoutburst interval}.\nocite{smws91}
A white dwarf that dominates the ultraviolet flux in quiescence
and cools after the outburst has been suggested as explanation
for the observed ultraviolet flux decrease.
This explanation is not compatible with the observations
in the ultraviolet of Z Cha, in which the contribution by white dwarf
and disk are determined separately. 
A cooling white dwarf doesn't explain our X-ray observations of YZ Cnc.

A well-known problem of the disk instability model is its failure
to describe the observation in short outbursts that the optical rise
precedes the ultraviolet rise by several hours (Pringle et al.\ 1986).
Various {\it ad hoc} suggestions have been made to explain this 
ultraviolet delay. These models suggest that the inner part of the disk 
continues to drain in quiescence. Such
models, which include the effect of a magnetic field of the white dwarf
(Livio \&\ Pringle 1992), a wind from the accretion disk in quiescence
(Meyer \&\ Meyer-Hofmeister 1994), and irradiation of the disk by
the (relatively) hot white dwarf (King  1997) possibly are compatible
with the decrease of the ultraviolet and X-ray flux during quiescence.

Finally, we note that the X-ray flux at the end of the quiescent interval
preceding the outburst peaking on JD\,2450912 is higher than the flux
measured during the two subsequent quiescent intervals.
This indicates variability in the flux level between different
outburst intervals, and shows that a trend in the quiescent
interval is best measured from a single interval.
The X-ray flux in the beginning of the superoutburst, on April 24,
is similar to the fluxes we measure in the beginning of the
ordinary outbursts.

This variability may be similar to the long-term variations that we
find by comparing our new observations with earlier observations
made with the ROSAT PSPC (see Sect.\ 4.1).
We do not find any clear correlation with the outburst pattern.
Thus, the ROSAT PSPC measurements were made in relatively long
quiescent intervals (11 days in Oct 1993, May 1994) and in a
relatively short quiescent interval (5 days in April 1991).
The ROSAT PSPC observations were obtained longer before the next
superoutburst than our new HRI observations; but the Einstein observation
was also made long before the next superoutburst.
The high countrate observed with the ROSAT PSPC therefore is not due
to a different length of the quiescent periods, nor to a different
location in the interval between superoutbursts.

\acknowledgements{PJW acknowledges support by PPARC as a postdoctoral fellow.}

\end{document}